\documentclass[12pt]{iopart}
\usepackage{iopams}
\usepackage{setstack}
\usepackage{graphicx}
\usepackage{bm}
\usepackage{amsopn}

\newcommand{\di}{\,\mathrm{d}}
\newcommand{\pdiff}[2]{\frac{\partial #1}{\partial #2}}
\newcommand{\diff}[2]{\frac{\di #1}{\di #2}}
\newcommand{\bv}{\bi{b}}
\newcommand{\cv}{\bi{c}}
\newcommand{\dv}{\bi{d}}
\newcommand{\ev}{\bi{e}}
\newcommand{\fv}{\bi{f}}
\newcommand{\gv}{\bi{g}}
\newcommand{\hv}{\bi{h}}
\newcommand{\mv}{\bi{m}}
\newcommand{\nv}{\bi{n}}
\newcommand{\pv}{\bi{p}}
\newcommand{\qv}{\bi{q}}
\newcommand{\rv}{\bi{r}}
\newcommand{\xvi}{\bi{x}_0}
\newcommand{\xv}{\bi{x}}
\newcommand{\zerov}{\boldsymbol{0}}
\newcommand{\Am}{\bi{A}}
\newcommand{\Em}{\bi{I}}
\newcommand{\Hm}{\bi{H}}
\newcommand{\Km}{\bi{K}}
\newcommand{\Qm}{\bi{Q}}
\newcommand{\Um}{\bi{U}}
\newcommand{\Vm}{\bi{V}}
\newcommand{\Zv}{\bi{Z}}
\newcommand{\im}{\mathrm{i}}
\newcommand{\real}{\mathbf{R}}
\DeclareMathOperator{\REAL}{Re}
\DeclareMathOperator{\IMAG}{Im}
\newcommand{\zetav}{{\bzeta}}
\newcommand{\chiv}{{\bchi}}
\newcommand{\psiv}{{\bpsi}}
\newcommand{\Sigmam}{{\bSigma}}
\newcommand{\Omegam}{{\bOmega}}
\newcommand{\inv}[1]{{#1}^{-1}}
\newcommand{\hconj}[1]{{#1}^{\dagger}}
\newcommand{\hypergeom}[4]{{_2F_1}\left(#1,#2;#3;#4\right)}

\begin{document}

\title[Collective motion of particles on a sphere]{Solvable model of the collective motion of heterogeneous particles interacting on a sphere}

\author{Takuma Tanaka}

\address{Department of Computational Intelligence and Systems Science, Interdisciplinary Graduate School of Science and Engineering, Tokyo Institute of Technology, Yokohama, Japan}
\ead{tanaka.takuma@gmail.com}
\begin{abstract}
% should not normally exceed 200 words
I propose a model of mutually interacting particles on an $M$-dimensional unit sphere.
I derive the dynamics of the particles by extending the dynamics of the Kuramoto--Sakaguchi model.
The dynamics include a natural-frequency matrix, which determines the motion of a particle with no external force, and an external force vector.
The position (state variable) of a particle at a given time is obtained by the projection transformation of the initial position of the particle.
The same projection transformation gives the position of the particles with the same natural-frequency matrix.
I show that the motion of the centre of mass of an infinite number of heterogeneous particles whose natural-frequency matrices are obtained from a class of multivariate Lorentz distribution is given by an $M$-dimensional ordinary differential equation in closed form.
This result is an extension of the Ott--Antonsen theory.
\end{abstract}

\submitto{\NJP} % or \JPA
% keywords: phase oscillator; synchronization; collective motion

\section{Introduction}
Systems consisting of many elements, such as a society, a school of fish, a flock of birds and the network of neurons in the brain, have attracted much research attention over years \cite{Hoppensteadt1997,Hayakawa2010,Bialek2012}.
Oscillation is a ubiquitous phenomenon found in systems such as these that consist of a large number of mutually interacting elements.
Other examples of these systems include reaction--diffusion systems \cite{Zhabotinsky1973}, electrochemical reactions \cite{Kiss2007}, electronic circuits \cite{Appleton1922,Nana2006}, spiking neurons \cite{Ermentrout1994}, the human gait \cite{Strogatz2005}, flashing of fireflies \cite{Buck1988}, the female menstrual cycle \cite{McClintock1971}, a linear array of CO$_2$ waveguide lasers \cite{Glova1996} and Josephson-junction arrays \cite{Tsang1991}.
Dynamics of oscillatory element $i$ can be described by
\[
\dot\xv_i = \chiv_i(\xv_i),
\]
where $\xv_i$ is the state vector and $\chiv_i(\xv_i)$ is the function defining the behaviour of element $i$.
The dynamics of the system consisting of these elements can be described by
\[
\dot\xv_i = \chiv_i(\xv_i)+\sum_{1\le j\le N}\psiv_{ij}(\xv_i,\xv_j),
\]
where $\psiv_{ij}(\xv_i,\xv_j)$ is the function characterising the interaction from $j$ to $i$ and $N$ is the number of oscillators in the system.
%The high number of degrees of freedom of elements and their heterogeneity make these systems so difficult to solve that sometimes one can only perform numerical simulations and observe the results.
In many cases, these systems are so difficult to solve that one can only perform numerical simulations and observe the results.
To gain insight into the collective phenomena of these complicated systems, we need to develop methods to analyse these systems.
The phase description or phase reduction of limit-cycle oscillators introduced by Kuramoto \cite{Kuramoto1984} is the most widely used method to analyse the synchronisation phenomena that lead to observed oscillations.
The phase description describes the state of a limit-cycle oscillator by a variable $\phi$ (called the phase) whose dynamics are
\[
\dot\phi = \omega+\Zv(\phi)\cdot\pv(t),
\]
where $\omega$ is the natural frequency, $\pv(t)$ is an external force and $\Zv(\phi)$ is the phase-sensitivity function.
If there is no external force, this system oscillates with constant frequency $\omega$.
The external force $\pv(t)$ advances or delays the phase.
$\Zv(t)$ determines the sensitivity of the oscillator to the external force.
If two oscillators with phases $\phi_1$ and $\phi_2$ are coupled to each other, the force exerted by oscillator 2 on oscillator 1 is given by the phase $\phi_2$, i.e. $\pv(\phi_2)$.
Assuming the coupling is weak, the long-time average of the mutual coupling $\Zv(\phi_1)\cdot\pv(\phi_2)$ can be regarded as a function of the phase difference $\phi_2-\phi_1$.
Thus, the dynamics of two oscillators are described by
\begin{eqnarray*}
\dot\phi_1 &=& \omega_1+\Gamma_{12}(\phi_1-\phi_2),\\
\dot\phi_2 &=& \omega_2+\Gamma_{21}(\phi_2-\phi_1),
\end{eqnarray*}
where $\Gamma_{ij}$ is the coupling function characterising the interaction from $j$ to $i$.

\begin{figure}
\begin{center}
\includegraphics[width=12cm]{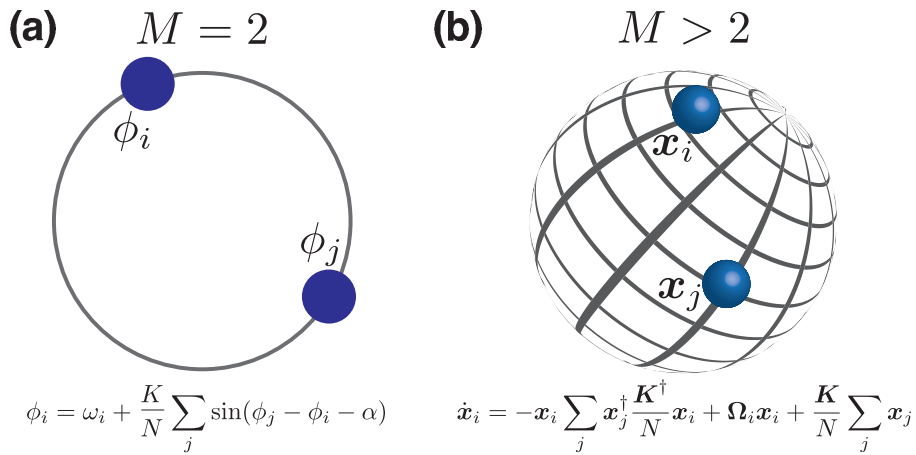}
\caption{\label{figSchema}
Kuramoto--Sakaguchi model (a) and proposed model (b).
The Kuramoto--Sakaguchi model may be regarded as the dynamical model of $N$ particles interacting on a circle.
The proposed model describes the dynamics of $N$ particles interacting on an $M$-dimensional sphere.
Here we set $\gv=\frac{\Km}{N}\sum_j\xv_j$.
}
\end{center}
\end{figure}

The Kuramoto--Sakaguchi model is the simplest model of weakly coupled oscillators and is described by
\[
\dot\phi_i = \omega_i+\frac{K}{N}\sum_{1\le j\le N}\sin(\phi_j-\phi_i-\alpha),
\]
where %$N$ is the number of oscillators in the system and
$K$ is the coupling strength \textbf{(\fref{figSchema} (a))}.
In the Kuramoto--Sakaguchi model, the coupling function is identical among all oscillator pairs.
In such a system, the natural frequency $\omega_i$ differs from oscillator to oscillator; the most intensively studied is the system in which the natural frequencies follow a Lorentz distribution.
The system has a critical coupling strength $K_c$ at which the synchronisation transition occurs.
If $K<K_c$, the system remains desynchronised and the order parameter $r$ (i.e. the centre of mass) defined by
\[
r = \frac{1}{N}\sum_{1\le i\le N}\exp(\im\phi_i)
\]
is zero.
For $K>K_c$, the system is synchronised and has non-zero $r$.
This model has been studied intensively and has contributed to the understanding of the synchronisation phenomena in general \cite{Acebron2005}.

Recently, two powerful methods to analyse the Kuramoto--Sakaguchi model have attracted attention.
First, Watanabe and Strogatz \cite{Watanabe1994} showed that the trajectory of a system of $N$-coupled oscillators is on three-dimensional invariant manifold if the system is described by
\begin{equation}
\dot\phi_i=\omega+\frac{K}{N}\sum_{1\le j\le N}\sin(\phi_j-\phi_i+\alpha)+J\sin(\phi_i+\beta), \label{WSdynamics}
\end{equation}
where $\omega$ is the natural frequency common to all oscillators; $K$, $J$, $\alpha$, and $\beta$ are time-dependent parameters.
The phase of a single oscillator in the system at time $t$ is determined by its initial phase and a nonlinear transformation involving three parameters.
An important point is that the transformation with the same parameters can be used to obtain the phase of any oscillator in the system.
Marvel and collaborators showed that the M\"obius transformation describes the $i$-th phase at time $t$ \cite{Goebel1995,Marvel2009}.
This transformation is a one-variable projection transformation in the complex plane
\begin{equation}
z_i(t)=\exp(\im\sigma)\frac{a-z_i(0)}{1-\bar a z_i(0)}, \label{Moebius}
\end{equation}
where $z_i(t)=\exp[\im\phi_i(t)]$ and both of $\sigma$ and $a$ are time-dependent and oscillator-independent parameters.
This type of M\"obius transformation maps the unit circle on the complex plane to itself.
Because the M\"obius transformation \eref{Moebius} has a real parameter $\sigma$ and a complex parameter $a$, all the trajectories of the system are confined to manifolds with three dimensions.
In other words, the dynamics of the system can be described by the dynamics of $\sigma$ and $a$.
This method simplifies the analysis of a system of oscillators with identical natural frequencies and enables a precise analysis of a wider range of synchronisation phenomena such as chimera states \cite{Laing2009} and persistent fluctuations in synchronisation rates \cite{Atsumi2012} than before.

Second, consider Kuramoto--Sakaguchi oscillators whose natural frequencies come from a Lorentz distribution
\[
p(\omega)=\frac{1}{\pi}\frac{\gamma}{\omega^2+\gamma^2},
\]
where $\gamma$ is the scale parameter.
Ott and Antonsen showed that the order parameter $r$ of an infinite number of such oscillators obeys the Stuart--Landau equation
\[
\dot r = \frac{1}{2}(K\cos\alpha-2\gamma-\im K\sin\alpha)r-\frac{K}{2}\exp(\im\alpha)|r|^2r
\]
under the condition that the initial phases are on a specified manifold \cite{Ott2008}.
Ott--Antonsen theory has been used to describe systems with bimodal natural-frequency distributions \cite{Martens2009} and systems with two subpopulations \cite{Pikovsky2011}.
In these studies, the systems contain an infinite number of heterogeneous oscillators, but are described by differential equations with a few dynamical variables, allowing the systems to be analysed.
Ott--Antonsen theory made the analysis of the synchronisation phenomena of complicated systems much simpler and much more thorough than previous methods.

The Kuramoto model with $\alpha=0$ and $\omega=0$ can be regarded as the XY spin model if noise is injected into the oscillators.
The Hamiltonian of the XY spin model is $H=-\sum_{\langle i,j\rangle}J_{ij}\cos(\phi_j-\phi_i)$, where the summation runs over all connected $i$ and $j$.
Spins in the XY spin model are two-dimensional unit vectors.
The XY spin model is generalised to the $n$-vector model, in which spins are unit vectors of arbitrary dimension.
The XY spin model and $n$-vector model have been used to analyse frustrated magnets.
The $n$-vector models are highly simplified and abstract models for describing magnetic spins; they do not necessarily correspond to physical substances.
However, solving the $n$-vector models has advanced our understanding of phase transition.
Moreover, the $n$-vector model with $n=0$ was suggested to correspond to self-avoiding walks \cite{deGennes1972,Bowers1973}.
Similarly, although percolation models in high-dimensional lattices and random graphs do not necessarily correspond to physical materials, the results of infinite-dimensional percolation were useful to interpret the results of percolation on complex networks such as the attack tolerance of networks \cite{Albert2000}.
The generalisation of such models allows us to predict what happens in novel problems and to find relationships between two seemingly unrelated physical systems.

The $n$-vector model, which is a high-dimensional extension of the XY spin model, has contributed to statistical physics by improving our understanding of phase transitions.
A high-dimensional extension of the Kuramoto model may similarly advance the understanding of the collective phenomena of coupled systems and can be applied to other fields.
If the extended model is solvable, it would have many possible applications.
In particular, if the methods similar to those of Watanabe and Strogatz and Ott and Antonsen are applicable to the model, it would be quite useful for understanding collective phenomena.

The study of collective phenomena goes back a long time.
In the 17th century, Huygens observed the antiphase locking of two pendulum clocks \cite{Pikovsky2001}; in the 18th century, Kaempfer reported the synchronised flashing of fireflies at the banks of the Chao Phraya \cite{Buck1988}.
Although these descriptions date back to the early modern period, the study of synchronisation phenomena advanced only in the latter half of the 20th century with the development of the phase-description method of limit-cycle oscillators \cite{Kuramoto1984,Winfree2001}.
Similarly, a solvable generalisation of the Kuramoto--Sakaguchi model would be useful for understanding the collective phenomena in a broader setting.

Motivated by this historical perspective, I propose an extension of the Kuramoto--Sakaguchi model to describe the collective motion of particles interacting on a unit sphere \textbf{(\fref{figSchema} (b))}.
First, by extending the framework of the Watanabe--Strogatz transformation, we derive the dynamics of the individual elements, which I call `particles'.
The present states of these particles, which I sometimes call `positions', are provided by the projection transformation (a high-dimensional extension of M\"obius transformation) of their initial values.
This dynamics of a particle are given by
\begin{eqnarray*}
\dot \xv &=& -(\hconj{\gv}\xv)\xv+\Omegam\xv+\gv\\
&=& -\xv\hconj{\gv}\xv+\Omegam\xv+\gv,
\end{eqnarray*}
where $\hconj{}$ denotes the Hermitian conjugate, $\xv$ is an $M$-dimensional real or complex vector representing the state of the particle, $\Omegam$ is an $M\times M$ matrix corresponding to the natural frequency of phase oscillators and $\gv$ determines the force exerted on the particle.
Assuming that $\xv$ is a two-dimensional real vector and assuming that $\xv$ is a one-dimensional complex vector lead to the same dynamics as the Kuramoto--Sakaguchi model.
The position of the particle at time $t$ is given by the same projection transformation, irrespective of the initial position $\xvi$.
Second, we derive the centre of mass $\rv$ of the particles that initially are uniformly distributed on the sphere.
The centre of mass corresponds to the order parameter of the Kuramoto--Sakaguchi model.
In particular, I show that the centre of mass is given by one of the parameter vectors of the projection transformation if the state vector $\xv$ is a complex vector.
Third, we extend the Ott--Antonsen theory to high-dimensional systems.
By assuming that the natural-frequency matrices of particles are obtained from a multivariate Lorentz distribution, I show that the centre of mass of particles with complex variables is described by low-dimensional ordinary differential equations.
In particular, I show that, if the particles in a system are attracted to $\Km\rv$, where $\Km$ is an $M\times M$ matrix, the dynamics of $\rv$ are described by
\[
\dot\rv = -\rv\hconj\rv\hconj\Km\rv
+\hat\Omegam\rv+\Km\rv,
\]
where $\hat\Omegam$ is a matrix determined by the probability distribution of the natural frequencies of the particles.
We derive the limit cycle from these dynamics and show that a transition emerges from a desynchronised state to a synchronised state that is similar to that found in the Kuramoto model.
I also show that this theoretically derived limit cycle agrees well with the results of numerical simulation.
Finally, I discuss the possible applications of the present model and problems to be addressed in the future.

\section{Results and discussion}

\subsection{Dynamics of the particles on the unit sphere induced by the projection transformation}

The M\"obius transformation \eref{Moebius} has been shown to underlie the dynamics of \eref{WSdynamics} \cite{Marvel2009}.
Marvel and collaborators showed that the phase of oscillator $i$ at time $t$ is given by the M\"obius transformation \eref{Moebius} of its phase at time $0$.
Notably, given the initial conditions of oscillators, the same parameters $\sigma$ and $a$ (i.e. the same M\"obius transformation) can be used to calculate the phases at $t$ of all oscillators in the system.
To extend the Kuramoto--Sakaguchi model to an $M$-dimensional system, we use the M\"obius transformation.
Because this transformation is a one-dimensional projection transformation, we derive the dynamics of variables whose present values are given by an $M$-dimensional projection transformation of the initial values.
We consider a system in which the $M$-dimensional state vector $\xv$ of a particle is given by the projection transformation
\begin{equation}
\xv = \frac{\Am\xvi+\bv}{\hconj\cv\xvi+d} \label{projection}
\end{equation}
 of the initial state $\xvi$, where the $M\times M$ matrix $\Am$, $M$-dimensional vectors $\bv$ and $\cv$ and scalar value $d$ are time dependent.
The system can be real valued or complex valued (i.e. $\xv$ can be a real or a complex vector).
To derive the dynamics of particles whose time evolution is described by \eref{projection}, we differentiate $\xv$ with respect to time to obtain
\begin{equation}
\dot\xv = \frac{(\dot\Am\xvi+\dot\bv)(\hconj\cv\xvi+d)-(\Am\xvi+\bv)(\hconj{\dot\cv}\xvi+\dot d)}{(\hconj\cv\xvi+d)^2}. \label{differentiation}
\end{equation}
Substituting
\begin{eqnarray*}
\xvi &=& \inv{(\Am-\xv\hconj\cv)}(d\xv-\bv)\\
&=& \frac{1}{1-\hconj\cv\inv\Am\xv}(\inv\Am-\hconj\cv\inv\Am\xv\inv\Am+\inv\Am\xv\hconj\cv\inv\Am)(d\xv-\bv)
\end{eqnarray*}
into $\Qm \xvi+\qv$, we have
\begin{eqnarray*}
\Qm \xvi+\qv
&=& \iota^{-1}[\Qm(\inv\Am-\hconj\cv\inv\Am\xv\inv\Am+\inv\Am\xv\hconj\cv\inv\Am)(d\xv-\bv)+\iota\qv]\\
&=& \iota^{-1}
[\Qm\inv\Am(d\xv-\bv)+\Qm(\hconj\cv\inv\Am\xv\inv\Am-\inv\Am\xv\hconj\cv\inv\Am)\bv+\iota\qv],
\end{eqnarray*}
where $\iota=1-\hconj\cv\inv\Am\xv$.
Replacing $\Qm$ and $\qv$ with $\hconj\cv$ and $d$, respectively, yields
\[
\hconj\cv\xvi+d = \kappa\iota^{-1},
\]
where $\kappa=d-\hconj\cv\inv\Am\bv$.
The first term in the numerator of \eref{differentiation} is
\[
\kappa\iota^{-2}
[\dot\Am\inv\Am(d\xv-\bv)+\dot\Am(\hconj\cv\inv\Am\xv\inv\Am-\inv\Am\xv\hconj\cv\inv\Am)\bv+\iota\dot\bv],
\]
and the second term in the numerator of \eref{differentiation} is
\[
\kappa\iota^{-2}
\xv
[\hconj{\dot\cv}\inv\Am(d\xv-\bv)+\hconj{\dot\cv}(\hconj\cv\inv\Am\xv\inv\Am-\inv\Am\xv\hconj\cv\inv\Am)\bv+\iota\dot d].
\]
Rearranging terms gives
\begin{eqnarray}
\dot\xv
&=& 
[\xv(-\kappa\hconj{\dot\cv}-\hconj{\dot\cv}\inv\Am\bv\hconj\cv+\dot d\hconj\cv)\inv\Am\xv \nonumber\\
&&
+(\kappa\dot\Am+\dot\Am\inv\Am\bv\hconj\cv-\dot\bv\hconj\cv+\hconj{\dot\cv}\inv\Am\bv\Am-\dot d\Am)\inv\Am\xv \nonumber\\
&&
-\dot\Am\inv\Am\bv+\dot\bv]
/\kappa.
\label{dzdt}
\end{eqnarray}
The dynamics
\begin{equation}
\dot\xv = \xv\hconj\hv\xv+\Omegam\xv+\gv \label{freedynamics}
\end{equation}
can be obtained from \eref{dzdt} if
\begin{eqnarray*}
\hconj\hv &=& \kappa^{-1}(-\kappa\hconj{\dot\cv}-\hconj{\dot\cv}\inv\Am\bv\hconj\cv+\dot d\hconj\cv)\inv\Am,\\
\Omegam &=& \kappa^{-1}(\kappa\dot\Am+\dot\Am\inv\Am\bv\hconj\cv-\dot\bv\hconj\cv+\hconj{\dot\cv}\inv\Am\bv\Am-\dot d\Am)\inv\Am,\\
\gv &=& \kappa^{-1}(-\dot\Am\inv\Am\bv+\dot\bv).
\end{eqnarray*}
To realise \eref{freedynamics}, the dynamics of the parameters of the projection transformation must be
\numparts
\begin{eqnarray}
\dot d &=& 0, \label{pdynamics1}\\
\hconj{\dot\cv} &=& \kappa\hconj\hv\Am\inv{(-\kappa\Em-\inv\Am\bv\hconj\cv)} \nonumber\\
&=& -\hconj\hv\Am\inv{(\Am+\kappa^{-1}\bv\hconj\cv)}\Am \nonumber\\
&=& -\hconj\hv\Am\left(\inv\Am-\frac{\kappa^{-1}\inv\Am\bv\hconj\cv\inv\Am}{1+\kappa^{-1}\hconj\cv\inv\Am\bv}\right)\Am \nonumber\\
&=& -\hconj\hv\Am(\inv\Am-d^{-1}\inv\Am\bv\hconj\cv\inv\Am)\Am \nonumber\\
&=& -\hconj\hv(\Am-d^{-1}\bv\hconj\cv)
,\\
\dot\Am &=& \kappa^{-1}(\kappa\Omegam\Am-\dot\Am\inv\Am\bv\hconj\cv+\dot\bv\hconj\cv-\hconj{\dot\cv}\inv\Am\bv\Am) \nonumber\\
&=& \kappa^{-1}[\kappa\Omegam\Am+(-\dot\Am\inv\Am\bv+\dot\bv)\hconj\cv-\hconj{\dot\cv}\inv\Am\bv\Am] \nonumber\\
&=& \kappa^{-1}(\kappa\Omegam\Am+\kappa\gv\hconj\cv
+\hconj\hv\bv\Am
-d^{-1}\hconj\hv\bv\hconj\cv\inv\Am\bv\Am) \nonumber\\
&=& \Omegam\Am+\gv\hconj\cv + d^{-1}\hconj\hv\bv\Am,\\
\dot\bv &=& \kappa\gv+\dot\Am\inv\Am\bv \nonumber\\
&=& \kappa\gv+
\Omegam\bv+\gv\hconj\cv\inv\Am\bv + d^{-1}\hconj\hv\bv\bv \nonumber\\
&=& d\gv+
\Omegam\bv + d^{-1}\hconj\hv\bv\bv. \label{pdynamics4}
\end{eqnarray}
\endnumparts
Thus, there are two ways to obtain the state $\xv$ at time $t$.
First, to directly calculate \eref{freedynamics} from the initial condition $\xvi$.
Second, to calculate the parameters of the transformation, $\Am$, $\bv$, $\cv$ and $d$, using \eref{pdynamics1}--\eref{pdynamics4} and the initial condition $\Am=\Em$, $\bv=\mathbf{0}$, $\cv=\mathbf{0}$ and $d=1$ and then map $\xvi$ to $\xv$ using \eref{projection}.
Note that, with these initial conditions for the parameters, $\Am$, $\bv$, $\cv$ and $d$, the relationship $\xv=\xvi$ is satisfied at $t=0$.
The first and the second method calculate the time evolution of the state vector and the transformation, respectively.
Thus, one can regard these methods as corresponding to the Sch\"odinger and Heisenberg representation of quantum mechanics.

The dynamics of phase oscillators can be regarded as those restricted to a unit circle in a plane.
Similarly, we restrict the dynamics of a particle to the sphere $|\xv|=1$.
The condition
\begin{eqnarray*}
\diff{|\xv|^2}{t}
&=& \hconj{\dot\xv}\xv+\hconj\xv\dot\xv\\
&=& (\hconj{\hv}+\hconj\gv)\xv+\hconj{\xv}(\Omegam+\hconj\Omegam)\xv+\hconj{\xv}(\hv+\gv)\\
&=& 0,
\end{eqnarray*}
is satisfied by the constraints
\numparts
\begin{eqnarray}
\gv &=& -\hv \label{assumptionone},\\
\Omegam &=& -\hconj\Omegam. \label{assumptiontwo}
\end{eqnarray}
\endnumparts
Thus, the dynamics are described by
\begin{equation}
\dot \xv = -\xv\hconj{\gv}\xv+\Omegam\xv+\gv, \label{dynamics}
\end{equation}
where $\gv$ is an arbitrary vector and $\Omegam$ is an antisymmetric matrix in the real-valued system and an anti-Hermitian matrix in the complex-valued system.
This is a subclass of the differential equations called the Riccati matrix differential equation \cite{Reid1972,Goebel1995}.

For a real two-dimensional vector 
\[
\xv=\left[\begin{array}{c}
x_1\\
x_2
\end{array}\right],
\]
\eref{dynamics} reduces to
\begin{eqnarray*}
\dot x_1 &=& -(g_1 x_1+g_2 x_2)x_1-\omega x_2+g_1,\\
\dot x_2 &=& -(g_1 x_1+g_2 x_2)x_2+\omega x_1+g_2,
\end{eqnarray*}
where we set 
\[
\gv=\left[\begin{array}{c}
g_1\\
g_2
\end{array}\right]
\]
and 
\[
\Omegam=\left(\begin{array}{cc}
0&-\omega\\
\omega&0
\end{array}\right).
\]
Replacing $x_1$ with $\cos\phi$ and $x_2$ with $\sin\phi$ gives
\begin{eqnarray*}
-\dot\phi\sin\phi &=& -(g_1 \cos\phi+g_2\sin\phi)\cos\phi-\omega\sin\phi+g_1\\
&=& -\sin\phi(\omega+g_2\cos\phi-g_1\sin\phi)\\
&=& -\sin\phi[\omega+K\sin(\psi-\phi)],
\end{eqnarray*}
where we assume $g_1=K\cos\psi$ and $g_2=K\sin\phi$, which is the dynamics of a Kuramoto oscillator under an external force.
Similarly, for a complex variable $\xv=[\exp(\im\phi)]$, \eref{dynamics} reduces to
\begin{eqnarray*}
\dot\phi &=& \im\bar{g}\exp(\im\phi)+\omega-\im g\exp(-\im\phi)\\
&=& \omega+2\IMAG [g\exp(-\im\phi)]\\
&=& \omega+K\sin(\psi-\phi),
\end{eqnarray*}
where $\IMAG$ denotes the imaginary part and we set $\Omegam=(\im\omega)$ and $\gv=[g]=[K\exp(\im\psi)/2]$,
which is the dynamics of a Kuramoto oscillator as well.
Thus, these two systems have the same dynamics.
However, in general, a system of $2M$-dimensional real state variables and a system of $M$-dimensional complex state variables do not have the same dynamics because the antisymmetric matrix $\Omegam$ has $2M^2-M$ real degrees of freedom in the former case, whereas the anti-Hermitian matrix $\Omegam$ has $M^2$ real degrees of freedom in the latter case.
These two examples suggest that $\gv$ and $\Omegam$ correspond to the external force and the natural frequency, respectively, in the Kuramoto--Sakaguchi model.
The natural-frequency matrix $\Omegam$ determines the direction and the speed of the rotation of the particle on the unit sphere.

Next, we consider whether the interaction terms $-\xv\hconj{\gv}\xv+\gv$ of \eref{dynamics} can be regarded as the tangential component of a central force on the surface of the unit sphere.
Let us assume that the central force is described by 
\[
f(|\gv-\xv|)\frac{\gv-\xv}{|\gv-\xv|},
\]
whose origin is $\gv$.
Assuming that $\hconj\xv\xv=1$, the tangential component is calculated by using the projection operator $\Em-\xv\hconj\xv$ as
\[
(\Em-\xv\hconj\xv)f(|\gv-\xv|)\frac{\gv-\xv}{|\gv-\xv|}
= f(|\gv-\xv|)\frac{\gv-\xv\hconj\xv\gv}{|\gv-\xv|},
\]
which gives $-\xv\hconj{\gv}\xv+\gv$ if $\xv$ and $\gv$ are real vectors and $f(|\gv-\xv|)=|\gv-\xv|$.
Thus, the interaction of particles in this system is the same as that for the system of objects that are attracted to $\gv$ by ideal springs and are restricted to lie on the $M$-dimensional unit sphere.

Potential energy is obtained by integrating the force $-\xv\hconj{\gv}\xv+\gv$.
Assuming that particle $i$ is attracted to particle $j$ (i.e. $\gv=\xv_j$) the force exerted on particle $i$ by particle $j$ is given by $-\xv_i\hconj{\xv}_j\xv_i+\xv_j$.
The potential energy is given by
\begin{eqnarray*}
E &=& -\int_{\xv_j}^{\xv_i}(-\xv\hconj{\xv}_j\xv+\xv_j)\cdot\di\xv\\
&=& -\int_{\xv_j}^{\xv_i}\xv_j\cdot\di\xv\\
&=& \xv_j\cdot\xv_j-\xv_j\cdot\xv_i\\
&=& -\xv_j\cdot\xv_i+\mathrm{const.},
\end{eqnarray*}
where we notice that the position vector $\xv$ on the sphere is orthogonal to the tangent vector $\di\xv$.
$E$ is the potential energy of the $n$-vector model.

\subsection{Dynamics of the parameters of the projection transformation}

Constrained by \eref{assumptionone} and \eref{assumptiontwo}, the dynamics of the parameters are
\numparts
\begin{eqnarray}
\dot\Am &=& \Omegam\Am+\gv\hconj\cv-\hconj\gv\bv\Am \label{dA},\\
\dot\bv
&=& -\bv\hconj\gv\bv+\Omegam\bv+\gv \label{db},\\
\hconj{\dot\cv} 
&=& \hconj\gv(\Am-\bv\hconj\cv),
\end{eqnarray}
\endnumparts
where we set $d=1$.
%The above differential equations are satisfied by the relation $\hconj\cv=\hconj\bv\Am$, which is, as we will see later, a constraint of the projection transformation preserving the unit sphere.
In this system, the parameters $\Am$, $\bv$ and $\cv$ of the projection transformation are dependent variables.
The projection transformation to and from the unit sphere satisfies
\[
\hconj{(\Am\xvi+\bv)}(\Am\xvi+\bv)=\hconj{(\hconj\cv\xvi+1)}(\hconj\cv\xvi+1)
\]
under the constraint $\hconj\xvi\xvi=1$, i.e.
\[
\hconj\xvi(\hconj\Am\Am-\cv\hconj\cv+|\bv|^2\Em-\Em)\xvi
+\hconj\xvi(\hconj\Am\bv-\cv)
+(\hconj\bv\Am-\hconj\cv)\xvi=0,
\]
therefore,
\numparts
\begin{eqnarray}
\hconj\cv &=& \hconj\bv\Am, \label{cbA}\\
\hconj\Am\Am &=& \cv\hconj\cv+(1-|\bv|^2)\Em \label{AA}.
\end{eqnarray}
\endnumparts
From \eref{cbA}, $\cv$ is determined by $\Am$ and $\bv$.
Substituting \eref{cbA} into \eref{AA}, we obtain
\[
\hconj\Am\Am = \hconj\Am\bv\hconj\bv\Am+(1-|\bv|^2)\Em.
\]
Rearranging the terms and multiplying by $\inv{\hconj\Am}$ from the left and by $\inv\Am$ from the right, we have
\[
\Em-\bv\hconj\bv=(1-|\bv|^2)\inv{(\Am\hconj\Am)}
\]
therefore,
\begin{eqnarray*}
\Am\hconj\Am &=& (1-|\bv|^2)\inv{(\Em-\bv\hconj\bv)}\\
&=& (1-|\bv|^2)\left(\Em+\frac{\bv\hconj\bv}{1-\hconj\bv\bv}\right)\\
&=& (1-|\bv|^2)\Em+\bv\hconj\bv.
\end{eqnarray*}
The matrix $\Am$ satisfying this equation is defined by
\[
\Am=\Hm^{1/2}\Um,
\]
where $\Um$ is an arbitrary orthogonal matrix if $\xv$ is a real vector and an arbitrary unitary matrix if $\xv$ is a complex vector and 
\[
\Hm^{1/2}=\Vm\Sigmam^{1/2}\hconj\Vm,
\]
with $\Sigmam^{1/2}=(\sqrt{\Sigma_{ij}})$ and $\Vm\Sigmam\hconj\Vm$ being the singular value decomposition of
\[
\Hm=(1-|\bv|^2)\Em+\bv\hconj\bv.
\]
$\Vm$ appears two times in the singular value decomposition of $\Hm$ because $\Hm$ is a Hermitian matrix.
Taken together, these relationships imply that the projection transformation to and from the unit sphere is determined by a vector $\bv$ and an orthogonal or unitary matrix $\Um$.
Because the degrees of freedom of an $M$-dimensional orthogonal matrix and an $M$-dimensional unitary matrix are $M(M-1)/2$ and $M^2$, respectively, the degree of freedom of the transformation is $M(M+1)/2$ in the real-valued system and $M^2+2M$ in the complex-valued system.
Thus, the real-valued systems with $M=2$ have three degrees of freedom, and the complex-valued systems with $M=1$ also have three degrees of freedom.
It is identical to the dimension of the invariant manifolds of Watanabe and Strogatz \cite{Watanabe1994}.
The vector $\bv$ is an eigenvector of $\Hm$ with eigenvalue 1 because
\begin{eqnarray*}
\Hm\bv &=& (1-|\bv|^2)\bv+\bv\hconj\bv\bv\\
&=& \bv.
\end{eqnarray*}
Vectors orthogonal to $\bv$ (i.e. vectors $\nv$ satisfying $\hconj\bv\nv=0$)
are eigenvectors of $\Hm$ with eigenvalues $1-|\bv|^2$ because
\begin{eqnarray*}
\Hm\nv &=& (1-|\bv|^2)\nv+\bv\hconj\bv\nv\\
&=& (1-|\bv|^2)\nv.
\end{eqnarray*}
From these observations, we obtain 
\begin{equation}
\Hm^{1/2}\bv=\bv \label{Hbv}
\end{equation}
and 
\begin{equation}
\Hm^{1/2}\nv=\sqrt{1-|\bv|^2}\nv. \label{Hnv}
\end{equation}
%\begin{equation}
%\Vm\Sigmam^{1/2}\hconj\Vm = \iota\Em+\gamma\bv\hconj\bv. \label{Asingular}
%\end{equation}
% and $\gamma=\frac{1-\sqrt{1-|\bv|^2}}{|\bv|^2}$.
$\Hm^{1/2}$ approaches $\Em$ in the limit $|\bv|\rightarrow 0$, which 
indicates that the initial conditions $\Am=\Em$ and $\bv=0$ are consistent with each other if we set $\Um=\Em$.

\subsection{Centre of mass of particles}

We set the initial conditions of the parameters of the transformation to $\Am=\Em$ and $\bv=\mathbf{0}$ to satisfy $\xv=\xvi$.
The projection transformation of $\xvi$ to $\xv$ is given as
\begin{eqnarray}
\xv &=& \frac{\Hm^{1/2}\Um\xvi+\bv}{\hconj\bv\Hm^{1/2}\Um\xvi+1} \nonumber\\
&=& \frac{\Hm^{1/2}\Um\xvi+\bv}{\hconj\bv\Um\xvi+1}. \label{projecttwo}
\end{eqnarray}
We now derive the centre of mass
\[
\mv = \frac{1}{N}\sum_{1\le i\le N} \xv_i
\]
of the particle.
We assume that, initially, the points are uniformly distributed on the unit sphere.
This assumption simplifies the calculation of the centre of mass and enables the low-dimensional description of the entire system.
Assuming that the points are uniformly distributed on the unit sphere in the initial conditions and that there are an infinite number of particles, we have
\[
\mv = \frac{1}{S_M}\int_{|\xvi|=1} \xv\di\xvi
\]
for real-valued systems and 
\[
\mv = \frac{1}{S_{2M}}\int_{|\xvi|=1} \xv\di\xvi
\]
for complex-valued systems,
where $S_M=\frac{2\pi^{M/2}}{\Gamma(M/2)}$ is the surface area of the $M$-dimensional unit sphere.
\Eref{projecttwo} suggests that the density of points and the centre of mass are completely determined by $\bv$ because $\Um$ does not affect the distribution after transformation if the particles are uniformly distributed on the unit sphere.
Thus, without loss of generality, we can replace $\Um$ of \eref{projecttwo} with $\Em$ for calculating the centre of mass $\mv$.
Decomposing $\xvi$ into 
\[
\xvi=\eta\bv_1+\nv,
\]
where $\bv_1=|\bv|^{-1}\bv$, $\eta=\hconj\bv_1\xvi$, and $\nv$ is a vector orthogonal to $\bv$ (i.e.  $\nv=\xvi-\eta\bv_1$) we have
\begin{eqnarray*}
\xv &=& \frac{\Hm^{1/2}(\eta\bv_1+\nv)+\bv}{\hconj\bv(\eta\bv_1+\nv)+1}\\
&=& \frac{(\eta|\bv|^{-1}+1)\bv+\sqrt{1-|\bv|^2}\nv}{\eta|\bv|+1},
\end{eqnarray*}
where we use \eref{Hbv} and \eref{Hnv} and set $\Um=\Em$.

%Note that $\eta^2+|\nv|^2=1$.

The centre of mass of the particles in real space is given by
\begin{eqnarray}
\mv &=& \frac{1}{S_M}\int_{|\xvi|=1} \frac{(\eta|\bv|^{-1}+1)\bv+\sqrt{1-|\bv|^2}\nv}{\eta|\bv|+1}\di\xvi \nonumber\\
&=& \frac{1}{S_M}\int_{-1}^{1}\int_{|\nv|=\sqrt{1-\eta^2}} \frac{(\eta|\bv|^{-1}+1)\bv+\sqrt{1-|\bv|^2}\nv}{\eta|\bv|+1} \di\nv\frac{1}{\sqrt{1-\eta^2}}\di\eta \nonumber\\
&=& \frac{S_{M-1}\bv}{S_M}
\int_{-1}^{1}\frac{\eta|\bv|^{-1}+1}{\eta|\bv|+1}\frac{(1-\eta^2)^{(M-2)/2}}{\sqrt{1-\eta^2}} \di\eta \nonumber\\
&=& \frac{S_{M-1}\bv}{S_M}
\int_{0}^{\pi}\frac{|\bv|^{-1}\cos\theta+1}{|\bv|\cos\theta+1}\sin^{M-2}\theta \di\theta \nonumber\\
&=& \frac{\Gamma(M/2)}{\sqrt{\pi}\Gamma[(M-1)/2]}\sqrt{\pi}\frac{\Gamma[(M+1)/2]}{\Gamma[(M+2)/2]}\hypergeom{1/2}{1}{(M+2)/2}{|\bv|^2}\bv \nonumber\\
&=& \frac{M-1}{M}\hypergeom{1/2}{1}{(M+2)/2}{|\bv|^2}\bv \label{realcenter}
,
\end{eqnarray}
where $\hypergeom{a}{b}{c}{x}$ is the ordinary hypergeometric function.

The centre of mass of the particles in complex space for $M=1$ is given by
\begin{eqnarray*}
\mv &=& \frac{1}{S_{2}}\int_{|\xvi|=1} \frac{(\eta|\bv|^{-1}+1)\bv+\sqrt{1-|\bv|^2}\nv}{\eta|\bv|+1}\di\xvi\\
&=& \frac{\bv}{2\pi}\int_{0}^{2\pi} \frac{|\bv|^{-1}\exp(\im\theta)+1}{|\bv|\exp(\im\theta)+1}\di\theta\\
&=& \bv,
\end{eqnarray*}
because $\nv=0$ in this case.
For $M\ge 2$, the centre of mass is given by
\begin{eqnarray*}
\mv &=& \frac{1}{S_{2M}}\int_{|\eta|\le 1}\int_{|\nv|=\sqrt{1-|\eta|^2}} \frac{(\eta|\bv|^{-1}+1)\bv+\sqrt{1-|\bv|^2}\nv}{\eta|\bv|+1} \di\nv\frac{1}{\sqrt{1-|\eta|^2}}\di\eta\\
&=& \frac{S_{2M-2}\bv}{S_{2M}}
\int_{|\eta|\le 1}\frac{\eta|\bv|^{-1}+1}{\eta|\bv|+1}\frac{(1-|\eta|^2)^{(2M-3)/2}}{\sqrt{1-|\eta|^2}} \di\eta\\
&=& \frac{S_{2M-2}\bv}{S_{2M}}
\int_0^1\int_{0}^{2\pi}\frac{|\bv|^{-1}r\exp(\im\theta)+1}{|\bv|r\exp(\im\theta)+1}(1-r^2)^{M-2}r\di\theta\di r\\
&=& \frac{S_{2M-2}\bv}{S_{2M}}
\int_0^1 2\pi(1-r^2)^{M-2}r\di r\\
&=& \frac{2\pi S_{2M-2}\bv}{2(M-1)S_{2M}}\\
&=& \bv.
\end{eqnarray*}
Thus, $\bv$ is the centre of mass of the system with complex variables.

\subsection{Results of simulation for the system with identical particles}

\begin{figure}
\begin{center}
\includegraphics[width=12cm]{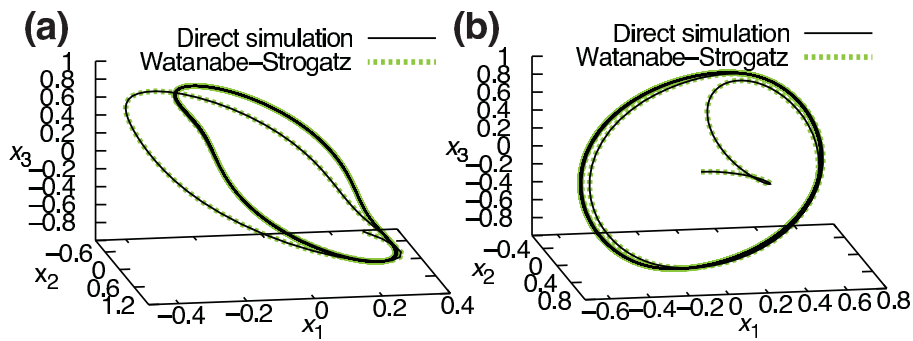}
\caption{\label{figWS}
Dynamics of the system of identical particles.
(a) Direct simulation of \eref{dynamics} (solid line) and the simulation by using \eref{dA} and \eref{db} of a particle under an external force (dashed line).
(b) Direct simulation (solid line) and approximation of the centre of mass (dashed line) for the system with $N=1000$ particles.
Simulation time is $T=300$ for both (a) and (b).
}
\end{center}
\end{figure}

\Fref{figWS} shows the results of simulation for particles described by a real three-dimensional variable.
In this system, all particles have the natural-frequency matrix
\begin{equation}
\Omegam=\left(\begin{array}{ccc}
0 & -1 & 1\\
1 & 0 & -1\\
-1 & 1 & 0
\end{array}\right),
\label{nfreqm}
\end{equation}
and experience the external force 
\[
\gv = 
\left[\begin{array}{c}
\cos t\\
\sin t\\
0
\end{array}\right].
\]
No mutual coupling among particles is introduced in \fref{figWS}(a).
The solid line in \fref{figWS}(a) is obtained by the direct simulation of \eref{dynamics} and shows the trajectory of one of the particles.
The dashed line shows the trajectory obtained by simulating the dynamics of parameters $\Am$ and $\bv$ by \eref{dA} and \eref{db} and transforming the initial conditions of the particle by the projection transformation \eref{projection} with these parameters.
These two simulation methods yield the same result.

Adding a mutual interaction and setting the external force to
\begin{equation}
\gv=\Km\mv+\left[\begin{array}{c}
\cos t\\
\sin t\\
0
\end{array}\right], \label{coupledg}
\end{equation}
where
$\mv$ is the centre of mass of $N=1000$ particles
and
\[
\Km=\frac{1}{2}\left(\begin{array}{ccc}
-1 & 0 & 1\\
0 & -1 & 0\\
1 & 0 & 1
\end{array}\right),
\]
I perform the direct simulation with $N=1000$ particles, which initially are randomly distributed on the unit sphere.
The same natural-frequency matrix as \eref{nfreqm} is used for all particles.
The solid line in \fref{figWS}(b) shows the trajectory of the centre of mass of the particles in the system.
Approximating $\mv$ of \eref{coupledg} by \eref{realcenter} (\Fref{figWS}(b), dashed line), I simulate the dynamics of parameters $\Am$ and $\bv$ by \eref{dA} and \eref{db} and obtain the trajectory with the projection transformation \eref{projection}.
\Fref{figWS}(b) shows that the direct simulation of the original system agrees well with this approximation.
This agreement occurs because the initial distribution of particles may be regarded as uniform when the number of particles is sufficiently large.

\subsection{Extension of the results of Ott and Antonsen}
Next, we consider a heterogeneous system in which the natural-frequency matrix $\Omegam$ varies from particle to particle.
Ott and Antonsen showed that the order parameter (i.e. the centre of mass) of an infinite number of Kuramoto--Sakaguchi oscillators obeys the Stuart--Landau equation if their natural frequencies are obtained from a Lorentz distribution \cite{Ott2008}.
To extend their results, we introduce two additional assumptions.
First, we assume a complex-valued system and that the state $\xv$ of the particle is a complex vector.
Because the centre of mass $\mv$ is approximated by $\bv$ in a complex-valued system, this assumption facilitates the derivation of the dynamics of the centre of mass.
Second, we assume that the natural-frequency matrix $\Omegam$ is drawn from a multivariate Lorentz distribution.
We assume that the entries of the natural-frequency matrix $\Omegam$ are given by the linear superposition of random variables drawn from Lorentz distributions:
\[
\Omegam = \Omegam_0+\sum_{1\le z\le Z}\zeta_z\Omegam_z,
\]
where $\zeta_z$ is obtained from the Lorentz distribution
with location parameter $\mu_z$ and scale parameter $\gamma_z>0$,
whose probability density is given by
\[
p_z(\zeta_z) = \frac{1}{\pi}\frac{\gamma_z}{(\zeta_z-\mu_z)^2+\gamma_z^2}.
\]
%Because $\Omegam_z\;(0\le z\le Z)$ must be an anti-Hermitian matrix,
%\[
%\Omegam_z = \hconj\Wm_z\Lambdam_z\Wm_z,
%\]
%where $\Wm$ is a unitary matrix and $\Lambdam$ is a diagonal matrix with purely imaginary elements.
Because $\Omegam$ must be an anti-Hermitian matrix, the eigenvalues of $\Omegam_z\;(0\le z\le Z)$ are purely imaginary.
We assume that 
\begin{eqnarray*}
\gv(t) &=& \Km(t)\int^{\real^Z}\mv(\Omegam,t)p(\zetav)\di \zetav+\fv(t)\\
&=& \Km(t)\rv(t)+\fv(t),
\end{eqnarray*}
where $\Km(t)$ is the coupling strength, $\fv(t)$ is the external force, and $p(\zetav) = \prod_{1\le z\le Z} p_z(\zeta_z)$.
Here, we show that, if $\Omegam_0$ is an anti-Hermitian matrix and the eigenvalues of $\Omegam_z$ are positive and imaginary for $z>0$, the dynamics of this heterogeneous system are described by
\begin{equation}
\dot\rv(t) = -\rv(t)\hconj{[\Km(t)\rv(t)+\fv(t)]}\rv(t)
+\hat\Omegam\rv(t)+\Km(t)\rv(t)+\fv(t),
\label{OA}
\end{equation}
where
\begin{equation}
\rv(t) = \mv(\hat\Omegam,t) \label{r}
\end{equation}
and
\[
\hat\Omegam = \Omegam_0+\sum_{1\le z\le Z}(\mu_z+\im\gamma_z)\Omegam_z.
\]

First, we derive the analytic continuation of $\mv(\Omegam,t)$, assuming that $\zeta_z$-s are complex numbers.
Because $\mv$ can be replaced by $\bv$ in complex-valued systems, we obtain from \eref{db} the dynamics of $\mv(\Omegam,t)$, which is
\[
\dot\mv(\Omegam,t) = -\mv(\Omegam,t)\hconj{\gv(t)}\mv(\Omegam,t)+\Omegam\mv(\Omegam,t)+\gv(t).
\]
Assuming that $\mv(\Omegam,t)$ satisfies the Cauchy--Riemann equations
\[
\pdiff{}{u_z}\mv(\Omegam,t) = -\im\pdiff{}{v_z}\mv(\Omegam,t),
\]
where $\zeta_z=u_z+\im v_z$ and $u_z$ and $v_z$ are real numbers, then $\dot\mv(\Omegam,t)$ satisfies the Cauchy--Riemann equations
\[
\pdiff{}{u_z}\dot\mv(\Omegam,t)=-\im\pdiff{}{v_z}\dot\mv(\Omegam,t)
\]
because
\begin{eqnarray*}
\pdiff{}{u_z}\diff{}{t}\mv(\Omegam,t) &=& \pdiff{}{u_z}\left(-\mv(\Omegam,t)\hconj{\gv(t)}\mv(\Omegam,t)+\Omegam\mv(\Omegam,t)+\gv(t)\right)\\
&=& -\pdiff{}{u_z}\mv(\Omegam,t)\hconj{\gv(t)}\mv(\Omegam,t)
-\mv(\Omegam,t)\hconj{\gv(t)}\pdiff{}{u_z}\mv(\Omegam,t)\\
&&+\pdiff{}{u_z}\Omegam\mv(\Omegam,t)
+\Omegam\pdiff{}{u_z}\mv(\Omegam,t)\\
&=& \im\pdiff{}{v_z}\mv(\Omegam,t)\hconj{\gv(t)}\mv(\Omegam,t)
+\im\mv(\Omegam,t)\hconj{\gv(t)}\pdiff{}{v_z}\mv(\Omegam,t)\\
&&-\im\pdiff{}{v_z}\Omegam\mv(\Omegam,t)
-\im\Omegam\pdiff{}{v_z}\mv(\Omegam,t)\\
&=& -\im\pdiff{}{v_z}\diff{}{t}\mv(\Omegam,t),
\end{eqnarray*}
where we used
\begin{eqnarray*}
\pdiff{}{u_z}\Omega_{ij}
&=& \pdiff{}{u_z}\left(\Omega_{0ij}+\sum_{1\le z\le Z}(u_z+\im v_z)\Omega_{zij}\right)\\
&=& \Omega_{zij}\\
&=& -\im\pdiff{}{v_z}\left(\Omega_{0ij}+\sum_{1\le z\le Z}(u_z+\im v_z)\Omega_{zij}\right)\\
&=& -\im\pdiff{}{v_z}\Omega_{ij}.
\end{eqnarray*}
These relations indicate that $\mv(\Omegam,t)$ is an analytic function of $\zeta_z$ if $\mv(\Omegam,t')$ is an analytic function of $\zeta_z$, where $t'<t$ \cite{Verhulst1996}.
Here, we have
\begin{eqnarray*}
\rv(t)
&=& \int^{\real^{Z-1}}\int^\real \mv(\Omegam,t)p_k(\zeta_k)\di\zeta_k\hat p_k(\hat\zetav_k)\di\hat\zetav_k\\
&=& \int^{\real^{Z-1}}\int^\real \mv(\Omegam,t)
\frac{1}{\pi}\frac{\gamma_k}{(\zeta_k-\mu_k)^2+\gamma_k^2}
\di\zeta_k
\hat p_k(\hat\zetav_k)\di\hat\zetav_k,
\end{eqnarray*}
where 
\begin{eqnarray*}
\hat\zetav_k &=& [\zeta_1,\,\ldots,\,\zeta_{k-1},\,\zeta_{k+1},\,\ldots,\,\zeta_z],\\
\hat p_k(\hat\zetav_k) &=& \prod_{1\le z\le Z,\;z\neq k} p_z(\zeta_z).
\end{eqnarray*}
Because $\mv(\Omegam,t)$ is an analytic function, we have
\begin{eqnarray}
&&\int^\real \mv(\Omegam,t)
\frac{1}{\pi}\frac{\gamma_k}{(\zeta_k-\mu_k)^2+\gamma_k^2}
\di\zeta_k \nonumber\\
&=&\lim_{S\rightarrow\infty}
\frac{1}{2\pi\im}\int_{-S}^{S} 
\frac{\mv(\Omegam,t)}{\zeta_k-\mu_k-\im\gamma_k}-\frac{\mv(\Omegam,t)}{\zeta_k-\mu_k+\im\gamma_k}
\di\zeta_k \nonumber\\
&& +
\lim_{S\rightarrow\infty}\frac{1}{2\pi\im}\int^{C}
\frac{\mv(\Omegam,t)}{\zeta_k-\mu_k-\im\gamma_k}-\frac{\mv(\Omegam,t)}{\zeta_k-\mu_k+\im\gamma_k}
\di\zeta_k
 \nonumber\\
&=&\mv(\hat\Omegam_k,t), \label{complexintegral}
\end{eqnarray}
where
$C$ is a semicircle in the upper half of the complex plane with radius $S$ and centred at the origin,
and
\[
\hat\Omegam_k = \Omegam_0+\Omegam_k(\mu_k+\im\gamma_k)+\sum_{1\le z\le Z,\;z\neq k}\zeta_z\Omegam_z,
\]
if the second integral on the right-hand side converges to zero.
Because $\Omegam_z$ is an anti-Hermitian matrix, the time evolution of $|\mv(\Omegam,t)|^2$ on $|\mv(\Omegam,t)|=1$ satisfies
\begin{eqnarray*}
\frac{1}{2}\diff{}{t}|\mv(\Omegam,t)|^2 &=& \REAL\left(\hconj{\mv(\Omegam,t)}\dot\mv(\Omegam,t)\right)\\
&=& \REAL\left(\hconj{\mv(\Omegam,t)}\Omegam\mv(\Omegam,t)\right)\\
&=& \REAL\left(\hconj{\mv(\Omegam,t)}\left(\Omegam_0+\sum_{1\le z\le Z}\zeta_z\Omegam_z\right)\mv(\Omegam,t)\right)\\
&=& \REAL\left(\hconj{\mv(\Omegam,t)}\sum_{1\le z\le Z}(u_z+\im v_z)\Omegam_z\mv(\Omegam,t)\right)\\
&=& \REAL\left(\hconj{\mv(\Omegam,t)}\sum_{1\le z\le Z}\im v_z\Omegam_z\mv(\Omegam,t)\right),
\end{eqnarray*}
where $\REAL$ denotes the real part.
If all eigenvalues of the Hermitian matrix $\im\Omegam_z$ are negative on the sphere,
\[
\diff{}{t}|\mv(\Omegam,t)|^2\le 0,
\]
then $|\mv(\Omegam,t)|=1$ when $\zeta_k$ is in the upper-half complex plane (i.e. $v_z\ge 0$).
Thus, $\mv(\Omega,t)$ remains finite if $\Omegam_z$ are anti-Hermitian matrices with positive imaginary eigenvalues and $\IMAG\zeta_k\ge 0$.
In the limit of large $S$, the dynamics of $\mv(\Omegam,t)$, where $|\zeta_k|=S$, can be approximated by
\[
\dot\mv(\Omegam,t) = \zeta_k\Omegam_k\mv(\Omegam,t),
\]
where we approximate $\Omegam$ by $\zeta_k\Omegam_k$ and ignore the terms without $\Omegam$.
If we use the same assumption that all eigenvalues of $\Omegam_k$ are positive and imaginary, then the real part of all eigenvalues of $\zeta_k\Omegam_k$ are negative under the condition $\IMAG\zeta_k>0$.
In this case, $\mv(\Omegam,t)$ approaches to zero as $S$ increases, and so the second integral of \eref{complexintegral} converges to zero in the limit of large $S$.
Thus, we have
\[
\rv(t)
= \int^{\real^{Z-1}}
\mv(\hat\Omegam_k,t)
\hat p_k(\hat\zetav_k)\di\hat\zetav_k.
\]
Because $\mv(\hat\Omegam_k,t)$ is an analytic function of $\zeta_z\;(z\neq k)$, the integration can be done recursively to obtain \eref{r}, so the dynamics of the order parameter $\rv(t)$ are given by \eref{OA}.
Note that $|\mv(\Omegam\,,t)|$ remains finite and converges to zero in the limit of large $S=|\zeta_z|\;(z\neq k)$ even if $\Omegam$ is replaced by $\hat\Omegam_k$.

\subsection{Limit-cycle oscillation of the centre of mass}
Assuming that there is no external force (i.e. $\fv(t)=0$) and that the mutual interactions among oscillators are constant  (i.e. $\Km(t)=\Km$), the dynamics are described by
\begin{equation}
\dot\rv(t) = -\rv(t)\hconj{\rv(t)}\hconj\Km\rv(t)
+\hat\Omegam\rv(t)+\Km\rv(t), \label{SL}
\end{equation}
which is a high-dimensional extension of the Stuart--Landau equation.
The stability of the desynchronised state, $\rv(t)=0$, is determined by the eigenvalues of $\hat\Omegam+\Km$.
%The eigenvalues of anti-Hermitian matrices are pure imaginary numbers, and the right eigenvectors equal the left eigenvectors; eigenvectors are orthogonal to each other.
Because $\Omegam_0$ is an anti-Hermitian matrix, $\Omegam_z\;(z>0)$ are anti-Hermitian matrices with positive imaginary eigenvalues and $\gamma_z>0$, so the real part of 
\[
\hconj\xv\hat\Omegam\xv = \hconj\xv\Omegam_0\xv+\sum_{1\le z\le Z}(\mu_z+\im\gamma_z)\hconj\xv\Omegam_z\xv
\]
is non-positive.
Therefore, all real parts of eigenvectors of $\hat\Omegam$ are non-positive, so the system without any mutual interaction ($\Km=0$) remains desynchronised.
%The dynamics of $\rv(t)$ are restricted to the subspace spanned by eigenvectors of $\hat\Omegam+\Km$ with eigenvalues with positive real components.
%In specific, 
If $\hat\Omegam+\Km$ has an eigenvector $\ev_i$ with eigenvalue $\lambda_i$, where $\REAL\lambda_i>0$, the limit cycle is given by 
\[
\rv_i=R_i\exp(\im\xi_i t)\ev_i,
\]
where
\begin{equation}
\im\xi_i = -R_i^2\hconj\ev_i\hconj\Km\ev_i+\lambda_i \label{limitcycle}
\end{equation}
and $|\ev_i|=1$.
Because 
\[
(\hat\Omegam+\Km)\ev_i=\lambda_i\ev_i,
\]
we get 
\[
\hconj\ev_i(\hconj{\hat\Omegam}+\hconj\Km)\ev_i=\bar\lambda_i.
\]
The real part of the right-hand side is positive because $\REAL\lambda_i>0$.
$\REAL(\hconj\ev_i\hconj{\hat\Omegam}\ev_i)\le 0$ because the real part of $\hconj\xv\hat\Omegam\xv$ is non-positive for any $\xv$.
Thus, we have $\REAL(\hconj\ev_i\hconj\Km\ev_i)>0$, and so \eref{limitcycle} is satisfied with a positive $R_i^2$ and a $\lambda_i$ with a positive real component.

%Assuming that $\REAL\lambda_1\ge\REAL\lambda_2\ge\cdots\ge\REAL\lambda_M$ and that $\REAL\lambda_1>0$,
Assuming that $\lambda_1$ is the eigenvalue with the largest real component,
we prove that only the limit cycle $\rv_1$ is stable.
Let us examine the time evolution of the perturbed solution
\[
\rv_i=(R_i+\rho)\exp(\im\xi_it+\im\theta)(\ev_i+\dv),
\]
where $\dv=\sum_{1\le j\le M,\;j\neq i}d_j\ev_j$, $|d_i|\ll 1$, $|\rho|\ll 1$ and $|\theta|\ll 1$.
Because we have
\begin{eqnarray*}
 \diff{\rv_i}{t}
&=& \dot\rho\exp(\im\xi_it+\im\theta)(\ev_i+\dv)
+(R_i+\rho)\exp(\im\xi_it+\im\theta)\dot\dv\\
&&+(R_i+\rho)\im(\xi_i+\dot\theta)\exp(\im\xi_it+\im\theta)(\ev_i+\dv)\\
&\approx& [\dot\rho+\im \dot\theta R_i+\im\rho\xi_i]\exp(\im\xi_it)\ev_i
+R_i\exp(\im\xi_it)\dot\dv\\
&&+\im R_i\xi_i\exp(\im\xi_it)\dv
+\im R_i \xi_i\exp(\im\xi_it+\im\theta)\ev_i
,
\end{eqnarray*}
\begin{eqnarray*}
&& -\rv_i\hconj\rv_i\hconj\Km\rv_i+\hat\Omegam\rv_i+\Km\rv_i\\
&=& -(R_i+\rho)^3\exp(\im\xi_i t+\im\theta)(\ev_i+\dv)\hconj{(\ev_i+\dv)}\hconj\Km(\ev_i+\dv)\\
&&+(R_i+\rho)\exp(\im\xi_i t+\im\theta)(\hat\Omegam+\Km)(\ev_i+\dv)\\
&\approx& [-3\rho R_i^2\ev_i\hconj\ev_i\hconj\Km\ev_i
-R_i^3\dv\hconj\ev_i\hconj\Km\ev_i
-R_i^3\ev_i\hconj\dv\hconj\Km\ev_i\\
&&\phantom{[}-R_i^3\ev_i\hconj\ev_i\hconj\Km\dv+\rho(\hat\Omegam+\Km)\ev_i
+R_i(\hat\Omegam+\Km) \dv]\exp(\im\xi_it)\\
&&+[-R_i^3\ev_i\hconj\ev_i\hconj\Km\ev_i+R_i(\hat\Omegam+\Km)\ev_i]\exp(\im\xi_it+\im\theta)\\
&=&
[-3\rho R_i^2\ev_i\hconj\ev_i\hconj\Km\ev_i
-R_i^3\dv\hconj\ev_i\hconj\Km\ev_i
-R_i^3\ev_i\hconj\dv\hconj\Km\ev_i\\
&&\phantom{[}-R_i^3\ev_i\hconj\ev_i\hconj\Km\dv+\rho(\hat\Omegam+\Km)\ev_i
+R_i(\hat\Omegam+\Km) \dv]\exp(\im\xi_it)\\
&&+\im R_i\xi_i\exp(\im\xi_it+\im\theta)\ev_i,
\end{eqnarray*}
replacing $\dv$ with $\sum_{1\le j\le M,\;j\neq i}d_j\ev_j$, we obtain
\begin{eqnarray*}
&&[\dot\rho+\im R_i\dot\theta+\im\xi_i\rho]\ev_i
+R_i\sum_{1\le j\le M,\;j\neq i}\dot d_j\ev_j
+\im R_i\xi_i\sum_{1\le j\le M,\;j\neq i}d_j\ev_j
\\
&=& 
3(\im\xi_i-\lambda_i)\rho\ev_i
+R_i(\im\xi_i-\lambda_i)\sum_{1\le j\le M,\;j\neq i}d_j\ev_j
-R_i^3\ev_i\sum_{1\le j\le M,\;j\neq i}\bar d_j\hconj\ev_j\hconj\Km\ev_i\\
&&-R_i^3\ev_i\hconj\ev_i\hconj\Km\sum_{1\le j\le M,\;j\neq i}d_j\ev_j
+\lambda_i\rho\ev_i
+R_i\sum_{1\le j\le M,\;j\neq i}\lambda_j d_j\ev_j,
\end{eqnarray*}
where we used $R_i^2\hconj\ev_i\hconj\Km\ev_i=\lambda_i-\im\xi_i$ and $(\hat\Omegam+\Km)\ev_i=\lambda\ev_i$.
Thus, the dynamics of $d_j$ is given by
\[
\dot d_j = (-\lambda_i+\lambda_j)d_j
\]
if $i\neq j$.
$d_1$ is unstable if $i>1$ because $\REAL\lambda_1>\REAL\lambda_i$.
Hence, the limit cycle $\rv_i$ is unstable if $i>1$.
Conversely, $d_j$ for $j>1$ is stable if $i=1$.
This fact allows us to set $\dv=\zerov$ in examining the stability of the limit cycle $\rv_1$.
The resulting dynamics of $\rho$ and $\theta$
\[
 \dot\rho+\im R_1\dot\theta
= 2(\im\xi_1-\lambda_1)\rho
\]
reduce to
\begin{eqnarray*}
\dot\rho &=& -2\REAL\lambda_1\rho,\\
R_1\dot\theta &=& 2(\xi_1-\IMAG\lambda_1)\rho.
\end{eqnarray*}
Hence, $\rho$ has a stable fixed-point at $\rho=0$, and $\theta$ is neutrally stable.

\subsection{Results of simulation of the heterogeneous particles with complex variables}
\begin{figure}
\begin{center}
\includegraphics[width=12cm]{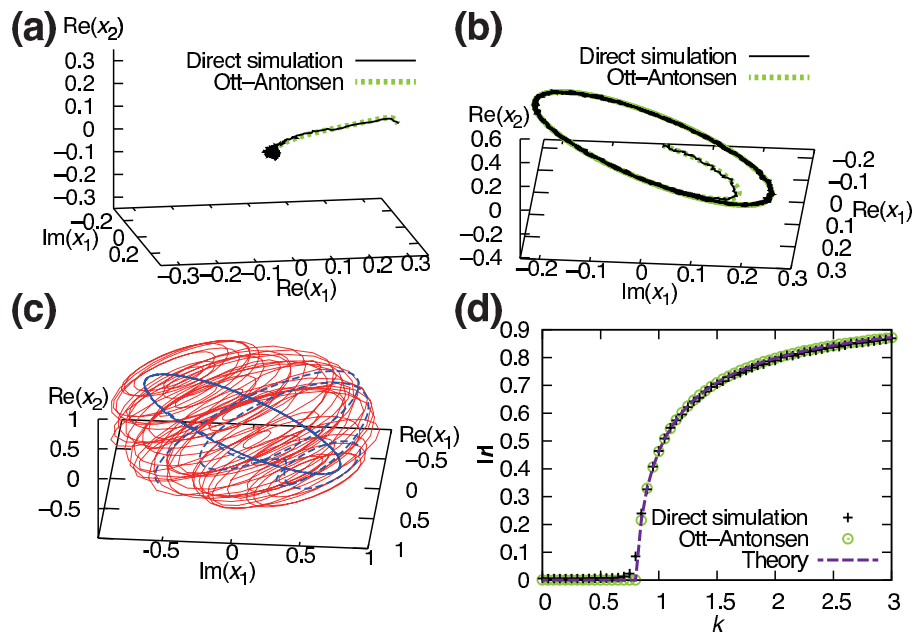}
\caption{\label{figOA}
Dynamics of the system of $N=10\,000$ heterogeneous particles.
(a,b) Direct simulations of the centre of mass (solid line) and low-dimensional dynamics (dashed line) for (a) $k=0.5$ and (b) $k=1$.
(c) Trajectory of two particles for $k=1$.
(d) Direct simulation of $|\rv|$ at $t=100$ (crosses), simulation of \eref{OA} at $t=100$ (circles) and theoretically derived radius of the limit cycle, $R_1$ (dashed line).
}
\end{center}
\end{figure}

\Fref{figOA}(a) and \fref{figOA}(b) compare the results of the direct simulation with the results of simulation of the reduced dynamics \eref{OA} for a system with $M=2$, $Z=1$:
$\Omegam_0=\left(\begin{array}{cc}
-\im & 1\\
-1 & -\im
\end{array}\right)$, 
$\Omegam_1=\left(\begin{array}{cc}
\im & 0\\
0 & \im
\end{array}\right)$, 
$\gamma_1=1$, $\mu_1=0$ and
$\Km=k\left(\begin{array}{cc}
-1 & -\im\\
\im & 1
\end{array}\right)$.
The solid lines are the trajectories of the order parameter $\rv$ of $N=10\,000$ particles obtained from the direct simulation of
\[
\dot\xv_i = -\xv_i\hconj\rv\hconj\Km\xv_i+\Omegam_i\xv_i+\Km\rv,
\]
where the order parameter is calculated by
\[
\rv = \frac{1}{N}\sum_{1\le i\le N}\xv_i.
\]
The dashed lines show the trajectories of $\rv$ obtained from the reduced dynamics \eref{SL}.
The results from the reduced dynamics agree quite well with those of the direct simulation.
By varying the value of $k$, I obtain the fixed point and the limit-cycle oscillation of $\rv$ (\fref{figOA}(a) and (b)).
This simple, low-dimensional behaviour of the limit cycle is in sharp contrast with the complicated motions of individual particles.
\Fref{figOA}(c) shows the trajectories of two particles in the system of \fref{figOA}(b).
One of the particles is entrained into the collective synchronisation whereas the other particle has a complicated trajectory.
\Fref{figOA}(d) shows how the radius of the limit cycle depends on $k$.
The radius obtained by solving \eref{limitcycle} (dashed line) agrees well with that obtained by the direct simulation (crosses) and by the reduced dynamics (circles).

\section{Conclusions}

In this study, we have extended the Kuramoto--Sakaguchi model to model particles interacting on a high-dimensional unit sphere.
The dynamics are described by a type of matrix Riccati differential equation, which is characterised by an external force vector and a natural-frequency matrix.
The position of a particle at a given time is obtained by the projection transformation of the initial position.
This result is an extension of the Watanabe--Strogatz theory.
By assuming that the particles are uniformly distributed on a unit sphere in the initial conditions, the centre of mass of the particles with the same natural-frequency matrix is determined by the vector of the parameters of the projection transformation.
In a system of particles with complex variables and natural-frequency matrices with a multivariate Lorentz distribution, the motion of the centre of mass of all the particles can be described by a high-dimensional extension of the Stuart--Landau equation.
This result is an extension of the Ott--Antonsen theory.
A periodic solution of the extended Stuart--Landau equation agrees with the motion of the centre of mass of the system.
Thus, we have shown that the collective motion of a system composed of elements with large degrees of freedom can be reduced to the dynamics of a low-dimensional system.

Although the present model is an extension to high degrees of freedom of the Kuramoto--Sakaguchi model, it differs from the previously proposed extensions \cite{Ritort1998,Gu2007}.
For example, Ritort introduced a variable into the model to keep the oscillators on the unit sphere \cite{Ritort1998}, whereas the particles of the present model (described by \eref{dynamics}) remain on the unit sphere without additional terms.
In addition, unlike the dynamic variables of the model by Gu and coworkers, which are represented by matrices \cite{Gu2007}, the dynamic variables in the present model are represented by vectors.
Finally, unlike the present model with heterogeneous particles, neither of these two models has been reported to be reducible to low-dimensional systems.

The present model provides a method to study new types of collective phenomena by reducing the system behaviour of elements with large degrees of freedom to that of a low-dimensional system.
The present model and its low-dimensional description can be applied to problems already studied from the viewpoint of Kuramoto oscillators \cite{Acebron2005}, such as time-delay systems \cite{Lee2009,Abrams2004}, systems of multiple-peak natural frequencies \cite{Martens2009}, non-local coupling \cite{Kuramoto2002}, dynamics on complex networks \cite{Boccaletti2006}, associative memory \cite{Aoyagi1995} and common-input synchronisation \cite{Teramae2004}.
In particular, because the system of oscillators with heterogeneous interaction delays \cite{Lee2009} and the system of oscillators whose natural frequencies obey a mixture of Lorentz distributions \cite{Martens2009} have been studied by using the Ott--Antonsen theory, I expect that the high-dimensional extension of these systems can be solved by the present method.

I have not found a physical system whose dynamics are described by the present model.
However, the phenomena observed in the present model might be useful in interpreting experimental observations of the systems that can be regarded as a population of particles on a sphere.
In particular, the normalised velocity of birds in a flock approximated by the Heisenberg model \cite{Bialek2012} could be analysed by using the real-valued systems with $M=3$.
% in which mutual interaction is approximated by ideal springs.
The results of the model suggest that the state transition from the desynchronised state to synchronised state can occur in systems with heterogeneous particles (\fref{figOA}(d)).
The present results also suggest that the trajectory of the centre of mass can be a very simple limit cycle (\fref{figOA}(b)) even if individual particles in the system exhibit complicated trajectories (\fref{figOA}(c)).
The present model could be used to approximate and to analyse systems exhibiting these properties.

\ack
This work was supported by MEXT/JSPS KAKENHI Grant Numbers 24651184 and 25115710.

\section*{References}
%\bibliographystyle{iopart-num}
%\bibliography{manuscript}

\begin{thebibliography}{10}
\expandafter\ifx\csname url\endcsname\relax
  \def\url#1{{\tt #1}}\fi
\expandafter\ifx\csname urlprefix\endcsname\relax\def\urlprefix{URL }\fi
\providecommand{\eprint}[2][]{\url{#2}}

\bibitem{Hoppensteadt1997}
Hoppensteadt F~C and Izhikevich E~M 1997 {\em {Weakly connected neural
  networks}\/} (Springer-Verlag)

\bibitem{Hayakawa2010}
Hayakawa Y 2010 {\em Europhys. Lett.\/} {\bf 89} 48004

\bibitem{Bialek2012}
Bialek W, Cavagna A, Giardina I, Mora T, Silvestri E, Viale M and Walczak A~M
  2012 {\em Proc. Natl. Acad. Sci.\/} {\bf 109} 4786--4791

\bibitem{Zhabotinsky1973}
Zhabotinsky A~M and Zaikin A~N 1973 {\em J. Theor. Biol.\/} {\bf 40} 45--61

\bibitem{Kiss2007}
Kiss I~Z, Rusin C~G, Kori H and Hudson J~L 2007 {\em Science\/} {\bf 316}
  1886--1889

\bibitem{Appleton1922}
Appleton E~V 1922 Automatic synchronization of triode oscillators {\em Proc.
  Cambridge Phil. Soc.\/} vol~21 pp 231--248

\bibitem{Nana2006}
Nana B and Woafo P 2006 {\em Physical Review E\/} {\bf 74} 46213

\bibitem{Ermentrout1994}
Ermentrout B 1994 {\em Neural Comput.\/} {\bf 6} 679--695

\bibitem{Strogatz2005}
Strogatz S~H {\em et~al.\/} 2005 {\em Nature\/} {\bf 438} 43--44

\bibitem{Buck1988}
Buck J 1988 {\em Q. Rev. Biol.\/} {\bf 63} 265--289

\bibitem{McClintock1971}
McClintock M~K 1971 {\em Nature\/} {\bf 229} 244--245

\bibitem{Glova1996}
Glova A~F, Kurchatov S~Y, Likhanskii V~V, Lysikov A~Y and Napartovich A~P 1996
  {\em Quantum Electronics\/} {\bf 26} 500--502

\bibitem{Tsang1991}
Tsang K~Y, Strogatz S~H and Wiesenfeld K 1991 {\em Phys. Rev. Lett.\/} {\bf 66}
  1094--1097

\bibitem{Kuramoto1984}
Kuramoto Y 1984 {\em {Chemical Oscillations, Waves, and Turbulence}\/}
  (Springer-Verlag)

\bibitem{Acebron2005}
Acebr{\'o}n J~A {\em et~al.\/} 2005 {\em Rev. Mod. Phys.\/} {\bf 77} 137--185

\bibitem{Watanabe1994}
Watanabe S and Strogatz S~H 1994 {\em Physica D\/} {\bf 74} 197--253

\bibitem{Goebel1995}
Goebel C~J 1995 {\em Physica D\/} {\bf 80} 18--20

\bibitem{Marvel2009}
Marvel S~A, Mirollo R~E and Strogatz S~H 2009 {\em Chaos\/} {\bf 19} 043104

\bibitem{Laing2009}
Laing C~R 2009 {\em Physica D\/} {\bf 238} 1569--1588

\bibitem{Atsumi2012}
Atsumi Y and Nakao H 2012 {\em Phys. Rev. E\/} {\bf 85}(5) 056207

\bibitem{Ott2008}
Ott E and Antonsen T~M 2008 {\em Chaos\/} {\bf 18} 037113

\bibitem{Martens2009}
Martens E~A, Barreto E, Strogatz S~H, Ott E, So P and Antonsen T~M 2009 {\em
  Phys. Rev. E\/} {\bf 79} 026204

\bibitem{Pikovsky2011}
Pikovsky A and Rosenblum M 2011 {\em Physica D\/} {\bf 240} 872--881

\bibitem{deGennes1972}
de~Gennes P~G 1972 {\em Phys. Lett. A\/} {\bf 38} 339--340

\bibitem{Bowers1973}
Bowers R~G and McKerrell A 1973 {\em J. Phys. C\/} {\bf 6} 2721

\bibitem{Albert2000}
Albert R, Jeong H and Barab{\'a}si A~L 2000 {\em Nature\/} {\bf 406} 378--382

\bibitem{Pikovsky2001}
Pikovsky A, Rosenblum M and Kurths J 2001 {\em Synchronization: a universal
  concept in nonlinear sciences\/} (Cambridge University Press)

\bibitem{Winfree2001}
Winfree A~T 2001 {\em {The geometry of biological time}\/} (Springer-Verlag)

\bibitem{Reid1972}
Reid W~T and Binkley C 1972 {\em Riccati Differential Equations\/} (Academic
  Press New York)

\bibitem{Verhulst1996}
Verhulst F 1996 {\em Nonlinear differential equations and dynamical systems\/}
  (Springer-Verlag)

\bibitem{Ritort1998}
Ritort F 1998 {\em Phys. Rev. Lett.\/} {\bf 80} 6--9

\bibitem{Gu2007}
Gu Z~M, Zhao M, Zhou T, Zhu C~P and Wang B~H 2007 {\em Phys. Lett. A\/} {\bf
  362} 115--119

\bibitem{Lee2009}
Lee W~S, Ott E and Antonsen T~M 2009 {\em Phys. Rev. Lett.\/} {\bf 103} 044101

\bibitem{Abrams2004}
Abrams D~M and Strogatz S~H 2004 {\em Phys. Rev. Lett.\/} {\bf 93} 174102

\bibitem{Kuramoto2002}
Kuramoto Y and Battogtokh D 2002 {\em Nonlinear Phenom. Complex Syst.\/} {\bf
  5} 380

\bibitem{Boccaletti2006}
Boccaletti S~S, Latora V, Moreno Y, Chavez M and Hwang D~U 2006 {\em Phys.
  Rep.\/} {\bf 424} 175--308

\bibitem{Aoyagi1995}
Aoyagi T 1995 {\em Phys. Rev. Lett.\/} {\bf 74} 4075--4078

\bibitem{Teramae2004}
Teramae J~N and Tanaka D 2004 {\em Phys. Rev. Lett.\/} {\bf 93} 204103

\end{thebibliography}
\providecommand{\newblock}{}

\end{document}